\documentclass[11pt]{article}

\usepackage[dvips]{graphicx}  

\begin{document}

\begin{center}
{\Large\bf Deriving the General Relativity Formalism: Understanding its Successes and Failures
\rule{0pt}{13pt}}\par

\bigskip

Reginald T. Cahill \\ 

{\small\it School of Chemistry, Physics and Earth Sciences, Flinders University,

Adelaide 5001, Australia\rule{0pt}{13pt}}\\

\raisebox{-1pt}{\footnotesize E-mail: Reg.Cahill@flinders.edu.au}\par

\vspace{2mm}

\bigskip\smallskip

{\small\parbox{11cm}{%
There are now at least eight experiments extending over more than 100 years that have detected the anisotropy of the speed of light, implying the absolute motion of the detecting apparatus through a dynamical space.   There are also many experiments that because of design flaws have failed to detect that anisotropy.  This light-speed anisotropy is consistent with relativistic effects and Lorentz symmetry, contrary to prevailing beliefs in physics.  The theoretical and experimental evidence implies that physics has failed to realise the existence of a dynamical 3-space, and that motion relative to that space is the cause of various relativistic effects, as proposed by Lorentz in 1899.  As well there is growing evidence that the phenomenon of gravity is more complex than previously believed, that Newtonian gravity appears to have failed even in the non-relativistic regime.  A new physics has emerged that builds upon this observed  dynamical 3-space and  provides a dynamical theory for that space.  This has resulted in a necessary generalisation of the Maxwell, Schr\"{o}dinger and Dirac equations, which then provide an explanation for gravity as an emergent phenomenon within the new physics.  From the generalised Dirac equation we show that the spacetime formalism is derivable, but as merely a mathematical construct whose geodesics arise from the trajectories of quantum wavepackets in the 3-space. However the metric of this spacetime  is shown not to satisfy the Hilbert-Einstein equations, except in the special case of the Schwarzschild metric.   Hence we demonstrate that the successes of the General Relativity formalism  have been more illusory than real, that its successes are in fact quite limited, which explains why it failed to account for the bore hole anomaly, the so-called `dark matter' spiral galaxy rotation anomaly, the systematics of black hole masses and so on. It also failed in that the dynamics of the 3-space is determined by two fundamental constants, namely $G$ and the fine structure constant $\alpha$.

\rule[0pt]{0pt}{0pt}}}\bigskip

\end{center}

\section{Introduction\label{section:introduction}}
We present here a derivation of  General Relativity \cite{Hilbert, Einstein}\footnote{There are ongoing developments regarding the priority issue of GR. However it is becoming clear that GR is actually a flawed theory of gravity, as discussed herein and elsewhere, which will put that issue into a different perspective.},  together with its geodesic  formalism, from a deeper theory in which the phenomenon of gravity  emerges as a quantum matter effect when coupled to a dynamical structured 3-space.  This derivation shows that both the flat and curved spacetime formalism arise as a purely mathematical construct - it has no ontological significance.  The experimental evidence is that a dynamical 3-space has been repeatedly observed over more than 100 years, and that physics has simply missed the existence of this 3-space [3-13].
 It is then necessary to  generalise the Maxwell, Schr\"{o}dinger and Dirac equations in order to couple these wave phenomena to the 3-space \cite{Schrod,newBH}.  In the case of the Schr\"{o}dinger and Dirac equations we see that gravity arises as an emergent quantum phenomena, while from the generalised Maxwell equations (as first suggested by Hertz in  1890 \cite{Hertz}) we obtain the light bending effects.  The experimental data has shown that the dynamics of the observed 3-space involves two fundamental constant $G$ and $\alpha$, the fine structure constant \cite{newBH,alpha,DM,galaxies,boreholes}.  This discovery is surely indicating that a new unified theory of reality is emerging.  Although the curved spacetime manifold formalism is shown to yield the correct matter and light trajectories, via the geodesic formalism, the metric of the spacetime does not satisfy the Hilbert-Einstein equations  \cite{Hilbert, Einstein} except in the special case of the Schwarzschild metric, and all but one of the putative successes of General Relativity involved that metric.   So the experimental, observational and theoretical evidence is that the successes of General Relativity were more illusory than real. It is also becoming clear that  it is the Lorentz interpretation \cite{Lorentz}  of relativistic effects that is being confirmed by experiment, namely that absolute motion of quantum and EM phenomena through the 3-space causes the well known relativistic effects.

It is commonly assumed that the successes of the special relativity formalism rest upon the assumption that the speed of light is the same for all observers, that the speed of light is isotropic for any observer, and that any claims of an observed anisotropy would be inconsistent with the successes of the special relativity formalism.  However this assumption is demonstrably wrong as discussed in \cite{anisotropy,Book}.  Of the eight experiments that have independently and consistently detected  that anisotropy the most recent \cite{anisotropy} has  also detected  gravitational waves, though not of the form supposedly predicted from General Relativity.

\section{Dynamics of Space\label{section:dynamics}}

At a deeper level an information-theoretic approach to modelling reality, {\it Process Physics} \cite{Book},  leads to an emergent structured `space'  which is 3-dimensional and dynamic, but where the 3-dimensionality is only approximate, in that if we ignore non-trivial topological aspects of the space, then it may be embedded in a 3-dimensional  geometrical manifold.  Here the space is a real existent discrete but fractal network of relationships or connectivities,  but the embedding space is purely a mathematical way of characterising the 3-dimensionality of the network.  This is illustrated in Fig.1. This is not an ether model; that notion involved a duality in that both the ether and the space in which it was embedded were both real.  Now the key point is that how we embed the network in the embedding space is very arbitrary: we could equally well rotate the embedding or use an embedding that has the network translated or translating.  These general requirements  then dictate the minimal dynamics for the actual network, at a phenomenological level.  To see this we assume  at a coarse grained level that the dynamical patterns within the network may be described by a velocity field ${\bf v}({\bf r},t)$, where ${\bf r}$ is the location of a small region in the network according to some arbitrary embedding.  For simplicity we assume here that the global topology of the network   is not significant for the local dynamics, and so we embed in an $E^3$, although a generalisation to an embedding in $S^3$ is straightforward.  The minimal dynamics then follows from the above by writing down the lowest-order zero-rank tensors, of dimension $1/t^2 $, that are invariant under translation and rotation, giving\footnote{Note that then, on dimensional grounds,  the  spatial dynamics cannot involve the speed of light $c$, except on the RHS where relativistic effects come into play if the speed of matter relative to the local space becomes large, see \cite{Book}. This has significant implications for the nature and   speed of  so-called `gravitational' waves.}

\begin{figure}[t]
\vspace{-3.5mm}\,\parbox{45.5mm}{\includegraphics[width=45.5mm]{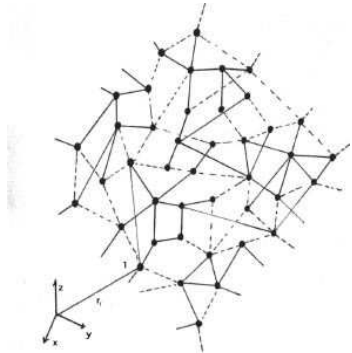}}
\parbox{90mm}{\caption{\small{  This is an iconic graph\-ic\-al representation of how a dynamical  network has its inherent approximate 3-dimen\-sion\-al\-ity displayed by an embedding in a math\-em\-at\-ical   space such as an $E^3$ or an $S^3$.  This space is not real; it is purely a mathematical artifact. Nevertheless this em\-bedd\-abi\-li\-ty helps de\-term\-ine the minimal dyn\-am\-ics for the network, as in (\ref{eqn:E1}).  At a deep\-er \,level \,the \,network \,is \,a \,quantum foam system \cite{Book}.  The dyn\-am\-ical space is not an ether model, as the em\-bedd\-ing space does not exist.}}}
\label{fig:Embedd}
\end{figure}

\begin{equation}
\nabla.\left(\frac{\partial {\bf v} }{\partial t}+({\bf v}.{\bf \nabla}){\bf v}\right)
+\frac{\alpha}{8}(tr D)^2 +\frac{\beta}{8}tr(D^2)=
-4\pi G\rho; \mbox{\ \ \ } D_{ij}=\frac{1}{2}\left(\frac{\partial v_i}{\partial x_j}+
\frac{\partial v_j}{\partial x_i}\right)
\label{eqn:E1}\end{equation}
where $\rho$ is the effective matter density.

In Process Physics  quantum matter  are topological defects in the network, but here it is sufficient to give a simple description in terms of an  effective density, but which can also model the `dark energy' effect and electromagnetic energy effects, which will be discussed elsewhere. We see that there are only four possible terms, and so we need at most three possible constants to describe the dynamics of space: $G, \alpha$ and $\beta$. $G$ will turn out  to be Newton's gravitational constant, and describes the rate of non-conservative flow of space into matter.  To determine the values of $\alpha$ and $\beta$ we must, at this stage, turn to experimental data.  

However most experimental data involving the dynamics of space is observed by detecting the so-called gravitational  acceleration of matter, although increasingly light bending is giving new information.  Now the acceleration ${\bf a}$ of the dynamical patterns in space is given by the Euler or convective expression
\begin{equation}
{\bf a}({\bf r},t)\equiv \lim_{\Delta t \rightarrow 0}\frac{{\bf v}({\bf r}+{\bf v}({\bf r},t)\Delta t,t+\Delta
t)-{\bf v}({\bf r},t)}{\Delta t} 
=\frac{\partial {\bf v}}{\partial t}+({\bf v}.\nabla ){\bf v}
\label{eqn:E3}\end{equation} 
and this appears in one of the terms in (\ref{eqn:E1}). As shown in \cite{Schrod} and discussed later herein the acceleration  ${\bf g}$ of quantum matter is identical to this acceleration, apart from vorticity and relativistic effects, and so the gravitational acceleration of matter is also given by (\ref{eqn:E3}).

Outside of a spherically symmetric distribution of matter,  of total mass $M$, we find that one solution of (\ref{eqn:E1}) is the velocity in-flow field  given by\footnote{To see that the flow is inward requires the modelling of the matter by essentially point-like particles. }
\begin{equation}
{\bf v}({\bf r})=-\hat{{\bf r}}\sqrt{\frac{2GM(1+\frac{\alpha}{2}+..)}{r}}
\label{eqn:E4}\end{equation}
but only when $\beta=-\alpha$,  for only then is the acceleration of matter, from (\ref{eqn:E3}), induced by this in-flow of the form
\begin{equation}
{\bf g}({\bf r})=-\hat{{\bf r}}\frac{GM(1+\frac{\alpha}{2}+..)}{r^2}
\label{eqn:E5}\end{equation}
 which  is Newton's Inverse Square Law of 1687, but with an effective  mass $M(1+\frac{\alpha}{2}+..)$ that is different from the actual mass $M$.  So Newton's law requires $\beta=-\alpha$ in (\ref{eqn:E1}) although at present a deeper explanation has not been found.  But we also see modifications coming from the 
$\alpha$-dependent terms.

In general because (\ref{eqn:E1}) is a scalar equation it is only applicable for vorticity-free flows $\nabla\times{\bf v}={\bf 0}$, for then we can write ${\bf v}=\nabla u$, and then (\ref{eqn:E1}) can always be solved to determine the time evolution of  $u({\bf r},t)$ given an initial form at some time  $t_0$.
The $\alpha$-dependent term in (\ref{eqn:E1})  (with now $\beta=-\alpha$) and the matter acceleration effect, now also given by (\ref{eqn:E3}),   permits   (\ref{eqn:E1})   to be written in the form
\begin{equation}
\nabla.{\bf g}=-4\pi G\rho-4\pi G \rho_{DM},
\label{eqn:E7}\end{equation}
where
\begin{equation}
\rho_{DM}({\bf r},t)\equiv\frac{\alpha}{32\pi G}( (tr D)^2-tr(D^2)),  
\label{eqn:E7b}\end{equation}
which  is an effective matter density that would be required to mimic the
 $\alpha$-dependent spatial self-interaction dynamics.
 Then (\ref{eqn:E7}) is the differential form for Newton's law of gravity but with an additional non-matter effective matter density. This effect explains the so-called `dark matter' effect in spiral galaxies, bore hole $g$ anomalies, and the systematics of galactic black hole massess. As shown elsewhere  it also explains, when used with the generalised Maxwell's equations,  the gravitational lensing of light by this `dark matter' effect.

 An intriguing aspect to the spatial dynamics is that it is non-local.  Historically this was first noticed by Newton who called it action-at-a-distance. To see this we can write  (\ref{eqn:E1}) as an integro-differential equation
 \begin{equation}
 \frac{\partial {\bf v}}{\partial t}=-\nabla\left(\frac{{\bf v}^2}{2}\right)+G\!\!\int d^3r^\prime
 \frac{\rho_{DM}({\bf r}^\prime, t)+\rho({\bf r}^\prime, t)}{|{\bf r}-{\bf r^\prime}|^3}({\bf r}-{\bf r^\prime})
 \label{eqn:E8}\end{equation}

 This shows a high degree of non-locality and non-linearity, and in particular that the behaviour of both $\rho_{DM}$ and $\rho$ manifest at a distance irrespective of the dynamics of the intervening space. This non-local behaviour is analogous to that in quantum systems. 
The non-local dynamics associated with the $\alpha$ dynamics has been tested in various situations, as discussed herein, and so its validity is well established.   This implies that the minimal spatial dynamics in  (\ref{eqn:E1})  involves non-local connectivities.

We term the dynamics of space in ({\ref{eqn:E1}) as a `flowing space'. This term can cause confusion because in normal language a `flow' implies movement of something relative to a background space; but here there is no existent background space, only the non-existent mathematical embedding space.
So here the `flow' refers to internal relative motion, that one parcel of space has a  motion relative to a nearby parcel of space.  Hence the absolute velocities in  ({\ref{eqn:E1})  have no observable meaning; that despite appearances it is only the relative velocities that have any dynamical significance.  Of course it is this requirement that determined the form of ({\ref{eqn:E1}),  as implemented via the embedding space technique.

 The new space dynamics shows that non-local effects are more general than just subtle effects in the quantum theory, for in the space dynamics this non-local dynamics is responsible for the supermassive black holes in galaxies. This non-local  dynamics is responsible for two other effects: (i) that the dynamics of space within an event horizon, say enclosing a black hole in-flow singularity,  affects the space outside of the horizon, even though EM radiation and matter cannot propagate out through the event horizon, as there the in-flow speed exceeds the speed of light.  So in this new physics we have the escape of information from within the event horizon,  and (ii) that  the universe overall is more highly connected than previously thought. This may explain why the universe is more uniform than expected on the basis of interactions limited by the speed of light, i.e we probably have a solution to the cosmological horizon problem.

However there is an additional role for the embedding space, namely as a coordinate system used by a set of cooperating observers. But again while this is useful for their discourse it is not real; it is not part of reality.

\section{Bore Holes and Black Holes\label{section:blackholes}}

A major recent discovery \cite{alpha, DM, galaxies, boreholes} has been that experimental data from the bore hole $g$ anomaly has revealed that $\alpha$ is the fine structure constant, to within experimental errors: $\alpha=e^2/\hbar c \approx 1/137.04$. This anomaly is that $g$ does not decrease as rapidly as predicted by Newtonian gravity or GR as we descend down a bore hole.  The dynamics in (\ref{eqn:E1}) and (\ref{eqn:E3}) gives
 the anomaly to be
 \begin{equation}
 \Delta g=2\pi\alpha G \rho d
 \label{eqn:E6}\end{equation}
where $d$ is the depth and $\rho$ is the density, being that of glacial ice in the case of the Greenland Ice Shelf experiments, or that of rock in the Nevada test site experiment. Clearly (\ref{eqn:E6})
permits the value of $\alpha$ to be determined from the data, giving  $\alpha=1/ (137.9 \pm 5)$ from the Greenland Ice Shelf data and, independently, $\alpha=1/ (136.8\pm 3)$ from the Nevada test site data \cite{boreholes}.

Eqn.(\ref{eqn:E1}) has `black hole' solutions.  The generic term `black hole' is used because they have a compact closed event horizon where the in-flow speed relative to the horizon equals the speed of light, but in other respects they differ from the putative black holes of General Relativity\footnote{It is probably the case that GR has no such solutions - they do not obey the boundary conditions at the singularity.} - in particular their gravitational acceleration is not inverse square law.  The evidence is that it is these new `black holes' from (\ref{eqn:E1}) that have been detected. There are two categories: (i) an in-flow singularity induced by the flow into a matter system, such as, herein, a spherical galaxy or globular cluster. These black holes are termed minimal black holes, as their effective mass is minimal, (ii) primordial naked black holes which then attract matter. These result in spiral galaxies, and the effective mass of the black hole is larger than required merely by the matter induced in-flow. These are therefore termed non-minimal black holes.   These explain the rapid formation of structure in the early universe, as the gravitational acceleration is approximately   $1/r$ rather than $1/r^2$. This is the feature that also explains the so-called `dark matter' effect in spiral galaxies.  

Consider the case where we have a spherically symmetric matter distribution at rest, on average with respect to distant space, and that the in-flow is time-independent and radially symmetric.  Then (\ref{eqn:E1})  can now be written in the form,  with $v^\prime=dv(r)/dr$, 
 \begin{equation}
2\frac{vv^\prime}{r} +(v^\prime)^2 + vv^{\prime\prime} =-4\pi G\rho(r)-4\pi G \rho_{DM}(v(r)), 
\label{eqn:Eradial}
\end{equation} 
or from (\ref{eqn:E8}) in the form
\begin{equation}\label{eqn:E9}
|{\bf v}({\bf r})|^2=2G\int d^3
r^\prime\frac{\rho_{DM}({\bf r}^\prime)+\rho({\bf r}^\prime)}{|{\bf r}-{\bf r}^\prime|}
\end{equation}
in which the angle integrations may be done to yield
\begin{equation}
v(r)^2=\frac{8\pi G}{r}\int_0^r s^2 \left[\rho_{DM}(s)+\rho(s)\right]ds 
 +8\pi G\int_r^\infty s
\left[\rho_{DM}(s)+\rho(s)\right]ds, 
\label{eqn:E9a}\end{equation}
 and now
 \begin{equation}
\rho_{DM}(r)= \frac{\alpha}{8\pi G}\left(\frac{v^2}{2r^2}+ \frac{vv^\prime}{r}\right).
\label{eqn:E10}\end{equation}
To obtain the induced minimal in-flow singularity to $O(\alpha)$ we substitute the non-$\alpha$ term in  (\ref{eqn:E9a}) into (\ref{eqn:E10}) giving the effective matter density that mimics the spatial self-interaction of the in-flow,
\begin{equation}
\rho_{DM}(r)=\frac{\alpha}{2r^2}\int_r^\infty s\rho(s)ds+O(\alpha^2).
\label{eqn:E9bb}\end{equation}
We see that  the effective `dark matter' effect is concentrated near the centre, and we find that the total
effective `dark matter' mass is
\begin{equation}
M_{DM} \equiv 4\pi\int_0^\infty r^2\rho_{DM}(r)dr  =\frac{4\pi\alpha}{2}\int_0^\infty
r^2\rho(r)dr+O(\alpha^2)  =\frac{\alpha}{2}M+O(\alpha^2). 
\label{eqn:E10b}\end{equation}
This result applies to any spherically symmetric matter distribution, and is the origin of the $\alpha$  terms in
(\ref{eqn:E4}) and (\ref{eqn:E5}).  It is thus responsible for the bore hole anomaly expression in (\ref{eqn:E6}).
This means that the bore hole anomaly is indicative of an in-flow singularity at the centre of the earth.  This contributes some  0.4\% of the effective mass of the earth, as defined by Newtonian gravity.  However in star systems this minimal black hole effect is more apparent, and we label $M_{DM}$ as $M_{BH}$.    So far black holes in 19  spherical star systems have been detected and together their masses give a best-fit value of $\alpha \approx 1/ 137.4$ \cite{newBH}.  

\section{Spiral Galaxy Rotation Anomaly\label{spiral}}
The application of the spatial dynamics to spiral galaxies is discussed in \cite{alpha, DM, galaxies, boreholes} where it is shown that a complete non-matter explanation of the spiral galaxy rotation speed anomaly is given: there is no such stuff as `dark matter' - it is an $\alpha$ determined spatial self-interaction effect.  Essentially even in the non-relativistic regime the Newtonian theory of gravity, with its `universal' Inverse Square Law, is deeply flawed.
Consider the non-perturbative solution of (\ref{eqn:E1}), say  for a galaxy with a non-spherical matter
distribution. Then numerical techniques are necessary, but beyond a sufficiently large distance  the in-flow will have spherical symmetry, and in that region we may use the spherically symmetric form of  (\ref{eqn:E1}).  Then (\ref{eqn:Eradial})   has an exact non-perturbative two-parameter class of analytic solutions
\begin{equation}
v(r) = K\left(\frac{1}{r}+\frac{1}{R}\left(\frac{R}{r}  \right)^{\displaystyle{\frac{\alpha}{2}}}  \right)^{1/2}
\label{eqn:vexact}\end{equation}
where $K$ and $R$ are arbitrary constants in the $\rho=0$ region, but whose values are determined by matching to the solution in the matter region. Here $R$ characterises the length scale of the non-perturbative part of this expression,  and $K$ depends on $\alpha$ and $G$ and details of the matter distribution.  The galactic circular orbital velocities of stars etc may be used to observe this in-flow process in a spiral galaxy and  from (\ref{eqn:E5}) and
(\ref{eqn:vexact}) we obtain a replacement for  the Newtonian  `inverse square law' ,
\begin{equation}
g(r)=\frac{K^2}{2} \left( \frac{1}{r^2}+\frac{\alpha}{2rR}\left(\frac{R}{r}\right)
^{\displaystyle{\frac{\alpha}{2}}} 
\right),
\label{eqn:gNewl}\end{equation}
in the asymptotic limit.     From  (\ref{eqn:gNewl}) the centripetal
acceleration  relation for circular orbits 
$v_O(r)=\sqrt{rg(r)}$  gives  a `universal rotation-speed curve'
\begin{equation}
v_O(r)=\frac{K}{2} \left( \frac{1}{r}+\frac{\alpha}{2R}\left(\frac{R}{r}\right)
^{\displaystyle{\frac{\alpha}{2}}} 
\right)^{1/2}
\label{eqn:vorbital}\end{equation}
 Because of the $\alpha$ dependent part this rotation-velocity curve  falls off extremely slowly with $r$, as is indeed observed for spiral galaxies. Of course it was the inability of the  Newtonian  and Einsteinian gravity theories to explain these observations that led to the  notion of `dark matter'.

\section{Generalising the Maxwell, Schr\"{o}dinger and Dirac Equations\label{section:generalising}}

One of the putative key tests of the GR formalism was the gravitational bending of light. This also immediately follows from the new space dynamics once we also generalise the Maxwell equations so that the electric and magnetic  fields are excitations of the dynamical space. The dynamics of the electric and magnetic fields  must then have the form, in empty space,  
\begin{equation}
\displaystyle{ \nabla \times {\bf E}=-\mu\left(\frac{\partial {\bf H}}{\partial t}+{\bf v.\nabla H}\right)};
 \mbox{\  }\displaystyle{ \nabla \times {\bf H}=\epsilon\left(\frac{\partial {\bf E}}{\partial t}+{\bf v.\nabla E}\right)} ;
 \mbox{\  }\displaystyle{\nabla.{\bf H}={\bf 0}}; \mbox{\ \ }
\displaystyle{\nabla.{\bf E}={\bf 0}}
\label{eqn:E18}\end{equation}
which was first suggested by Hertz in 1890  \cite{Hertz}. As easily determined  the speed of EM radiation is now $c=1/\sqrt{\mu\epsilon}$ with respect to the space, and in general not with respect to the observer if the observer is moving through space, as experiment has indicated again and again.
In particular the in-flow in (\ref{eqn:E4}) causes a refraction effect of light passing close to the sun, with the angle of deflection given by
\begin{equation}
\delta=2\frac{v^2}{c^2}=\frac{4GM(1+\frac{\alpha}{2}+..)}{c^2d}
\label{eqn:E19}\end{equation}
where $v$ is the in-flow speed at distance $d$  and $d$ is the impact parameter, here the radius of the sun. Hence the  observed deflection of $8.4\times10^{-6}$ radians is actually a measure of the in-flow speed at the sun's surface, and that gives $v=615$km/s.  At the earth distance the sun induced spatial  in-flow speed is 42km/s, and this has been extracted from the 1925/26 gas-mode interferometer Miller data \cite{AMGE,Book}. These radial in-flows are to be vectorially summed to the galactic flow of some 400km/s, but since that flow is much more uniform it does not affect the light bending by the sun in-flow component. The vector superposition effect for spatial flows is only approximate, and is discussed  in \cite{Book}. The solar system has a galactic velocity of some 400$\pm$30km/s in the direction RA=5.2hr, Dec=-67$^0$, as confirmed in a new light-speed anisotropy experiment \cite{anisotropy}. Hence the deflection of light by the sun is a way of directly measuring the spatial in-flow speed at the sun's surface, and has nothing to do with an actual curved spacetime. These generalised Maxwell equations also predict gravitational lensing produced by the large in-flows associated with new `black holes' in galaxies.  So again this effect permits the direct observation of the these  black hole effects with their non inverse-square-law accelerations.

Next consider the generalised Schr\"{o}dinger equation  \cite{Schrod}
\begin{equation}
i\hbar\frac{\partial  \psi({\bf r},t)}{\partial t}=H(t)\psi({\bf r},t),
\label{eqn:equiv7}\end{equation}
where the free-fall hamiltonian is
\begin{equation}
H(t)=-i\hbar\left({\bf
v}.\nabla+\frac{1}{2}\nabla.{\bf v}\right)-\frac{\hbar^2}{2m}\nabla^2
\label{eqn:equiv8}\end{equation}
As discussed in \cite{Schrod} this is uniquely defined by the requirement that  the wave function be attached to the dynamical space, and not to the embedding space, and that $H(t)$ be hermitian. We can compute the acceleration of a localised wave packet  according to
\begin{equation}{\bf g}\equiv\frac{d^2}{dt^2}\left(\psi(t),{\bf r}\psi(t)\right)  
=\frac{\partial{\bf v}}{\partial t}+({\bf v}.\nabla){\bf v}+
(\nabla\times{\bf v})\times{\bf v}_R
\label{eqn:E11}\end{equation}
where ${\bf v}_R={\bf v}_0-{\bf v}$  is the velocity of the wave packet relative to the local space, as ${\bf v}_0$ is  the velocity relative to the embedding space. Apart from the vorticity term which causes rotation of the wave packet\footnote{This explains the Lense-Thirring effect, and such vorticity  is being detected by the Gravity Probe B satellite gyroscope experiment\cite{GPB}.} we see, as promised, that this matter acceleration is equal to that of the space itself, as in (\ref{eqn:E3}). This is the first derivation of the phenomenon of gravity from a deeper theory: gravity is a quantum effect - namely the refraction of quantum waves by the internal differential motion of the substructure  patterns to space itself. Note that the equivalence principle has now been explained, as this `gravitational' acceleration is independent of the mass $m$ of the quantum system. 

An analogous generalisation of the Dirac equation is also necessary giving the coupling of the spinor to the actual dynamical space, and again not to the embedding space as has been the case up until now, 
\begin{equation}
i\hbar\frac{\partial \psi}{\partial t}=-i\hbar\left(  c{\vec{ \alpha.}}\nabla + {\bf
v}.\nabla+\frac{1}{2}\nabla.{\bf v}  \right)\psi+\beta m c^2\psi
\label{eqn:12}\end{equation}
where $\vec{\alpha}$ and $\beta$ are the usual Dirac matrices. Repeating the analysis in (\ref{eqn:E11}) for the space-induced acceleration we obtain
\begin{equation}\label{eqn:E12}
{\bf g}=\displaystyle{\frac{\partial {\bf v}}{\partial t}}+({\bf v}.{\bf \nabla}){\bf
v}+({\bf \nabla}\times{\bf v})\times{\bf v}_R-\frac{{\bf
v}_R}{1-\displaystyle{\frac{{\bf v}_R^2}{c^2}}}
\frac{1}{2}\frac{d}{dt}\left(\frac{{\bf v}_R^2}{c^2}\right)
\label{eqn:E13a}\end{equation}
which generalises  (\ref{eqn:E11}) by having a term which limits the speed of the wave packet relative to space to be $<\!c$. This equation specifies the trajectory of a spinor wave packet in the dynamical space.

\section{The Spacetime and Geodesic Formalism\label{section:spacetime}}

The curved spacetime explanation for gravity is widely known. Here an explanation for its putative success is given, for there is a natural definition of a spacetime  that arises from (\ref{eqn:E1}), but that it is purely a mathematical construction with no ontological status - it is a mere mathematical artifact. 
We shall now show how this leads to both the spacetime mathematical construct and that the geodesic for matter worldlines in that spacetime is equivalent  to trajectories from  (\ref{eqn:E12}).  First we note that (\ref{eqn:E12}) may be obtained by extremising the time-dilated elapsed time 
\begin{equation}
\tau[{\bf r}_0]=\int dt \left(1-\frac{{\bf v}_R^2}{c^2}\right)^{1/2}
\label{eqn:E13}\end{equation}  
with respect to the particle trajectory ${\bf r}_0(t)$ \cite{Book}. This happens because of the Fermat least-time effect for waves: only along the minimal time trajectory do the quantum waves  remain in phase under small variations of the path. This again emphasises  that gravity is a quantum effect.   We now introduce a spacetime mathematical construct according to the metric
\begin{equation}
ds^2=dt^2 -(d{\bf r}-{\bf v}({\bf r},t)dt)^2/c^2 
=g_{\mu\nu}dx^{\mu}dx^\nu
\label{eqn:E14}\end{equation}
Then according to this metric the elapsed time in (\ref{eqn:E13}) is
\begin{equation}
\tau=\int dt\sqrt{g_{\mu\nu}\frac{dx^{\mu}}{dt}\frac{dx^{\nu}}{dt}},
\label{eqn:E14b}\end{equation}
and the minimisation of  (\ref{eqn:E14b}) leads to the geodesics of the spacetime, which are thus equivalent to the trajectories from (\ref{eqn:E13}), namely (\ref{eqn:E13a}).
Hence by coupling the Dirac spinor dynamics to the space dynamics we derive the geodesic formalism of General Relativity as a quantum effect, but without reference to the Hilbert-Einstein equations for the induced metric.  Indeed in general the metric of  this induced spacetime will not satisfy  these equations as the dynamical space involves the $\alpha$-dependent  dynamics, and $\alpha$ is missing from GR.   
So why did GR appear to succeed in a number of key tests where the Schwarzschild metric was used?  The answer is provided by identifying the induced spacetime metric corresponding to the in-flow in (\ref{eqn:E4}) outside of a spherical matter system, such as the earth.  Then (\ref{eqn:E14})  becomes
 \begin{equation}
ds^2=dt^{ 2}-\frac{1}{c^2}(dr+\sqrt{\frac{2GM(1+\frac{\alpha}{2}+..)}{r}}dt)^2
-\frac{1}{c^2}r^2(d\theta^{ 2}+\sin^2(\theta)d\phi^2).
\label{eqn:E15}\end{equation}
 Making the change of variables\footnote{No unique choice of variables is required. This choice simply leads to a well-known form for the metric.}
$t\rightarrow t^\prime$ and
$\bf{r}\rightarrow {\bf r}^\prime= {\bf r}$ with
\begin{equation}
t^\prime=t-
\frac{2}{c}\sqrt{\frac{2 GM(1{+}\frac{\alpha}{2}{+}\dots)r}{c^2}}
+\frac{4\ GM(1{+}\frac{\alpha}{2}{+}\dots)}{c^3}\,\mbox{tanh}^{-1}
\sqrt{\frac{2 GM(1{+}
\frac{\alpha}{2}{+}\dots)}{c^2r}}
\label{eqn:E16}\end{equation}
this becomes (and now dropping the prime notation)
\begin{equation}
ds^2=\left(1-\frac{2GM(1+\frac{\alpha}{2}+..)}{c^2r}\right)dt^{ 2} 
-\frac{1}{c^2}r^{ 2}(d\theta^2+\sin^2(\theta)d\phi^2) 
-\frac{dr^{ 2}}{c^2\left(1-{\displaystyle\frac{
2GM(1+\frac{\alpha}{2}+..)}{ c^2r}}\right)}
\label{eqn:E17}\end{equation}
which is  one form of the the Schwarzschild metric but with the $\alpha$-dynamics induced effective mass shift. Of course this is only valid outside of the spherical matter distribution, as that is the proviso also on (\ref{eqn:E4}). As well the above particular change of coordinates also introduces spurious singularities at the event horizon\footnote{The event horizon of  (\ref{eqn:E17}) is at a different radius from the actual event horizon of the black hole solutions that arise from   (\ref{eqn:E1})},  but other choices do not do this. 
Hence in the case of the Schwarzschild metric the dynamics missing from both the Newtonian theory of gravity and General Relativity is merely hidden in a mass redefinition, and so didn't affect the various standard tests of GR, or even of Newtonian gravity.  Note that as well we see that the Schwarzschild metric is none other than Newtonian gravity in disguise, except for the mass shift.  While we have now explained why the GR formalism appeared to work, it is also clear that this formalism hides the manifest dynamics of the dynamical space, and which has also been directly detected in gas-mode interferometer and coaxial-cable experiments.

\section{Hilbert-Einstein Equations\label{section:hilberteinstein}}

Here we show that the metric of the spacetime  manifold   (\ref{eqn:E14}) emerging from (\ref{eqn:E1}) via  the  generalised Dirac equation  satisfies the 
Hilbert-Einstein GR \cite{Hilbert,Einstein} equations, but {\it only} in the limit $\alpha\rightarrow 0$. The GR equations are
\begin{equation}
G_{\mu\nu}\equiv R_{\mu\nu}-\frac{1}{2}Rg_{\mu\nu}=\frac{8\pi G}{c^2} T_{\mu\nu},
\label{eqn:32}\end{equation}
where  $G_{\mu\nu}$ is  the Einstein tensor, $T_{\mu\nu}$ is the  energy-momentum tensor,
$R_{\mu\nu}=R^\alpha_{\mu\alpha\nu}$ and
$R=g^{\mu\nu}R_{\mu\nu}$ and
$g^{\mu\nu}$ is the matrix inverse of $g_{\mu\nu}$. The curvature tensor is
\begin{equation}
R^\rho_{\mu\sigma\nu}=\Gamma^\rho_{\mu\nu,\sigma}-\Gamma^\rho_{\mu\sigma,\nu}+
\Gamma^\rho_{\alpha\sigma}\Gamma^\alpha_{\mu\nu}-\Gamma^\rho_{\alpha\nu}\Gamma^\alpha_{\mu\sigma},
\label{eqn:curvature}\end{equation}
where $\Gamma^\alpha_{\mu\sigma}$ is the affine connection
\begin{equation}
\Gamma^\alpha_{\mu\sigma}=\frac{1}{2} g^{\alpha\nu}\left(\frac{\partial g_{\nu\mu}}{\partial x^\sigma}+
\frac{\partial g_{\nu\sigma}}{\partial x^\mu}-\frac{\partial g_{\mu\sigma}}{\partial x^\nu} \right).
\label{eqn:affine}\end{equation}
Let us  substitute the metric in (\ref{eqn:E14}) into  (\ref{eqn:32}) using (\ref{eqn:curvature}) and (\ref{eqn:affine}).   The various components of the
Einstein tensor are then found to be
\begin{eqnarray}\label{eqn:G}
G_{00}&=&\sum_{i,j=1,2,3}v_i\mathcal{G}_{ij}
v_j-c^2\sum_{j=1,2,3}\mathcal{G}_{0j}v_j-c^2\sum_{i=1,2,3}v_i\mathcal{G}_{i0}+c^2\mathcal{G}_{00}, 
\nonumber\\ G_{i0}&=&-\sum_{j=1,2,3}\mathcal{G}_{ij}v_j+c^2\mathcal{G}_{i0},   \mbox{ \ \ \ \ } i=1,2,3.
\nonumber\\ G_{ij}&=&\mathcal{G}_{ij},   \mbox{ \ \ \ \ } i,j=1,2,3.
\end{eqnarray}
where the  $\mathcal{G}_{\mu\nu}$ are  given by
\begin{eqnarray}\label{eqn:GT}
\mathcal{G}_{00}&=&\frac{1}{2}((trD)^2-tr(D^2)), \nonumber\\
\mathcal{G}_{i0}&=&\mathcal{G}_{0i}=-\frac{1}{2}(\nabla\times(\nabla\times{\bf v}))_i,   \mbox{ \ \ \ \ }
i=1,2,3.\nonumber\\ 
\mathcal{G}_{ij}&=&
\frac{d}{dt}(D_{ij}-\delta_{ij}trD)+(D_{ij}-\frac{1}{2}\delta_{ij}trD)trD\nonumber\\ & &
-\frac{1}{2}\delta_{ij}tr(D^2)+(\Omega D-D\Omega)_{ij},  \mbox{ \ \ \ \ } i,j=1,2,3.
\end{eqnarray}
In vacuum, with $T_{\mu\nu}=0$, we find from (\ref{eqn:32}) and (\ref{eqn:G}) that $G_{\mu\nu}=0$ implies that  
$\mathcal{G}_{\mu\nu}=0$. 
We see the the Hilbert-Einstein equations demand that 
\begin{equation}
((trD)^2-tr(D^2))=0.
\label{eqg:DMGR}\end{equation}  
but it is these terms in   (\ref{eqn:E1})  that explain the various gravitational anomalies. This simply corresponds to the fact that GR does not permit the `dark matter' effect  according to (\ref{eqn:E7b}), and this happens because GR was forced to agree with Newtonian
gravity, in the appropriate limits, and that theory also has no such effect. As well in GR the energy-momentum
tensor
$T_{\mu\nu}$ is not permitted to make any reference to absolute linear motion of the matter; only  the relative
motion of matter or absolute rotational motion is permitted, contrary to the experiments [3-13].

It is very significant to note that the above exposition of the GR formalism for  the  metric in (\ref{eqn:E14}) is exact. Then taking the trace of the $\mathcal{G}_{ij}$ equation in (\ref{eqn:GT}) we obtain,
also exactly, and in the case of zero vorticity, and outside of matter so that
$T_{\mu\nu}=0$,
\begin{equation}
\frac{\partial }{\partial t}(\nabla.{\bf v})+\nabla.(({\bf
v}.{\bf \nabla}){\bf v})=0
\label{eqn:f3vacuum}\end{equation}
which is the Newtonian `velocity field' formulation of Newtonian gravity outside of matter, as in (\ref{eqn:E1}) but with $\alpha=\beta=0$.  

\section{Conclusions}

The extensive {\it non-null} experimental evidence over more than 100 years for the anisotropy of the one-way speed of light  has finally been understood with the startling conclusion that a dynamical structured 3-space exists and is responsible for the various special relativity effects and even Lorentz symmetry.  As reviewed herein this has lead to an explanation for the phenomenon of gravity, that it is a quantum matter effect arising from the refraction of the matter quantum waves in the dynamical 3-space.  In deriving the dynamical theory for this 3-space it has been discovered from various experimental and observational data that its self-interaction coupling constant in none other than the fine structure constant $\alpha$.  From the generalised Dirac equation we have shown that the quantum matter trajectories in the 3-space may be determined from the geodesics of a curved spacetime manifold, but that this spacetime has no ontological status - it is purely a mathematical construct.  We have also been able to derive the Hilbert-Einstein General Relativity formalism, but that derivation shows that it is only valid in the special limit of $\alpha\rightarrow 0$. For the case of the external Schwarzschild metric the $\alpha$-dynamics amounts to a rescaling of the mass, and so had no observational consequences - that is why the various tests of GR using that metric appeared to be successful. Hence only by finally being able to explain gravity, and by being able to derive the spacetime formalism, do we discover the reasons for its successes and failures. One notable failure of the Hilbert-Einstein theory of gravity was its inability to account for the so-called `dark matter' effect.   Not discussed here is the explanation for the success of the Global Positioning System (GPS) and that the dynamical equations for the 3-space have cosmological Hubble expansion solutions.

This work is supported by an Australian Research Council Discovery Grant 2005-2006:   {\it Development and Study of a New Theory of Gravity}.

\end{document}